\documentclass[9pt,lineno,dvipsnames,sort]{elife_redux}  

\usepackage{url}                  	
\usepackage{doi}           	 		
\usepackage{lipsum}                 
\usepackage[version=4]{mhchem}
\usepackage{siunitx}
\usepackage{gensymb}
\DeclareSIUnit\Molar{M}
\usepackage{float}
\usepackage{dsfont}

\usepackage{bbold}
\usepackage{amsmath}

\usepackage[algoruled,noend]{algorithm2e}
\usepackage{listings}
\usepackage{fancyvrb}
\fvset{fontsize=\small}

\title{Teaching deep neural networks to localize single molecules for super-resolution microscopy}

\author[1,2,3]{Artur Speiser} 
\author[4,5]{Lucas-Raphael M{\"u}ller}
\author[4]{Ulf Matti}
\author[8]{Christopher J. Obara}
\author[9,10]{Wesley R. Legant}
\author[4]{Jonas Ries}
\author[1,2,6,7,*]{Jakob H.\ Macke}
\author[8,*]{Srinivas C.\ Turaga}

\affil[1]{Computational Neuroengineering,
            Department of Electrical and Computer Engineering,
            Technical University of Munich, Munich, Germany}
\affil[2]{research center caesar, an associate of the Max Planck Society,
            Bonn, Germany}
\affil[3]{International Max Planck Research School `Brain and Behavior',
            Bonn/Florida}
\affil[4]{European Molecular Biology Laboratory, Heidelberg, Germany}
\affil[5]{Ruprecht Karl University of Heidelberg, Heidelberg, Germany}
\affil[6]{Excellence Cluster Machine Learning, T\"ubingen University, Germany} 
\affil[7]{Max Planck Institute for Intelligent Systems, T\"ubingen, Germany} 
\affil[8]{HHMI Janelia Research Campus, Ashburn VA, USA}
\affil[9]{Department of Pharmacology, University of North Carolina, Chapel Hill, USA}
\affil[10]{Department of Biomedical Engineering, University of North Carolina, Chapel Hill, USA}

\corr{Artur.Speiser@tum.de}{}

\contrib[*]{These authors contributed equally to this work}

\newcommand{\beginsupplement}{%
        \setcounter{table}{0}
        \renewcommand{\thetable}{S\arabic{table}}%
        \setcounter{figure}{0}
        \renewcommand{\thefigure}{S\arabic{figure}}%
     }

\begin{document}

\maketitle


\begin{abstract} 
Single-molecule localization fluorescence microscopy constructs super-resolution images by sequential imaging and computational localization of sparsely activated fluorophores. 
Accurate and efficient fluorophore localization algorithms are key to the success of this computational microscopy method.
We present a novel localization algorithm based on deep learning which significantly improves upon the state of the art. 
Our contributions are a novel network architecture for simultaneous detection and localization, and new loss function which phrases detection and localization as a Bayesian inference problem, and thus allows the network to provide uncertainty-estimates. In contrast to standard methods which independently process imaging frames, our network architecture uses temporal context from multiple sequentially imaged frames to detect and localize molecules. 
We demonstrate the power of our method across a variety of datasets, imaging modalities, signal to noise ratios, and fluorophore densities.  While existing localization algorithms can achieve optimal localization accuracy at low fluorophore densities, they are confounded by high densities. Our method is the first deep-learning based approach which achieves state-of-the-art on the SMLM2016 challenge. It achieves the best scores on 12 out of 12 data-sets when comparing both detection accuracy and precision, and excels at high densities. Finally, we investigate how unsupervised learning can be used to make the network robust against mismatch between simulated and real data.  The lessons learned here are more generally relevant for the training of deep networks to solve challenging Bayesian inverse problems on spatially extended domains in biology and physics.
\end{abstract}


\newpage 

\section{Introduction} 

\begin{figure}
\includegraphics[width=1.0\textwidth, trim={0.5cm 6.0cm 0.5cm 0.5cm},clip]{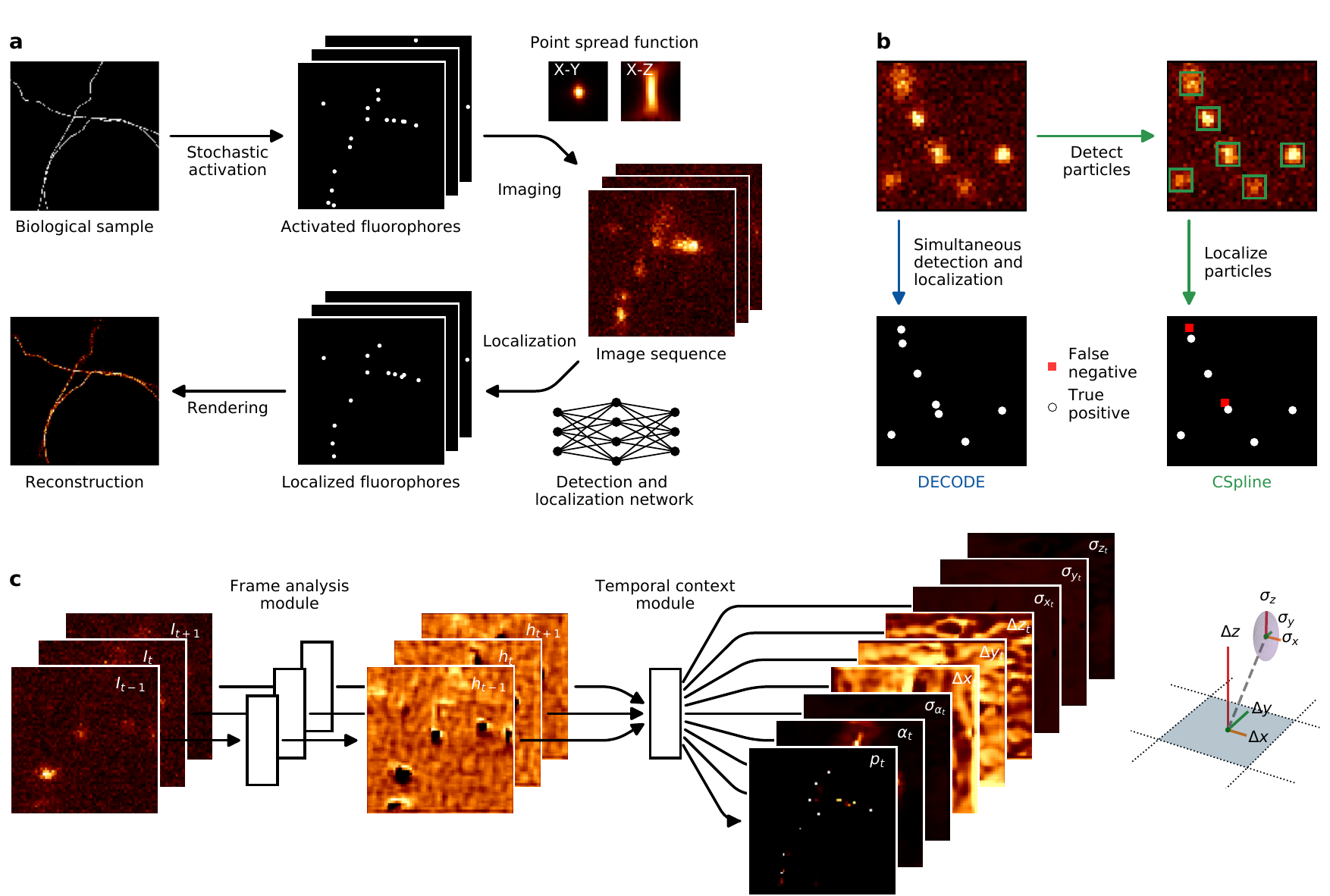}
\caption{
{\bf Source reconstruction for Single-Molecule Localization Microscopy (SMLM):}
 {\bf a)} Fluorophores are stochastically activated and recorded using fluorescence microscopy. A localization algorithm infers the underlying sources from noisy and blurred imaging measurements. Rendering methods turn inferred sources into an estimate of the underlying structure.
  {\bf b)} Classical image-processing algorithms for SMLM source localization (such as CSpline \cite{Spliner}) are based on a two-step approach (detect/localize), whereas our approach (DECODE) uses a  neural network for simultaneous detection and localization. 
{\bf c)} DECODE network for simultaneous detection and localization of fluorophores. Hidden features are extracted from each consecutive imaging frame by the first stage of the network by the \emph{frame analysis module}. These frame specific features are integrated by a \emph{temporal context module} leading to a prediction of 8 output maps:  a binary map of fluorophore detections $p_t$, a map predicting the brightness of the corresponding detected fluorophore $\alpha_t$,  three maps of the three spatial coordinates of the detected fluorophore , relative to the to the center of the detected pixel, $\Delta x_t,\Delta y_t,\Delta z_t$, and three maps of the associated uncertainties (standard deviations) $\sigma_{x_t},\sigma_{y_t},\sigma_{z_t}$.
\label{fig:concept}}
\end{figure}

Super-resolution microscopy techniques such as stochastic optical reconstruction microscopy (STORM) \cite{storm} and photo-activated localization microscopy (PALM) \cite{palm} have made it possible to observe biological structures and processes that where not accessible through optical microscopy due to the 
 Abbe diffraction limit. 
These techniques, commonly referred to as Single Molecule Localization Microscopy (SMLM), critically rely on computational methods for accurately localizing sparsely activated fluorophores \cite{review_deschout} (Fig. \ref{fig:concept}a).
  State-of-the-art localization algorithms typically operate in two steps: first, single fluorophore candidates are detected and extracted from the images, and second, fluorophores are localized by fitting a high resolution ``generative'' model of the point-spread function (PSF) to the image. To deal with overlapping fluorophores, peaks are either rejected based on a statistical test for the presence of multiple fluorophores (single emitter fitting \cite{smap,rapidstorm,localizer}), or emitters are added throughout the fitting procedure until a predetermined threshold for the goodness of fit is met (multi-emitter fitting  \cite{Spliner, daostorm,thunderstorm}). 
More recently, deep learning approaches have been used to perform the localization step \cite{dl_smlm1,dl_smlm2}.

This general approach can be highly effective under favourable conditions of high SNR and low fluorophore density \cite{rieger_review}. However, even multi-emitter approaches produce sub-par results in datasets with high fluorophore densities.  As was noted in a systematic comparison of multiple algorithms on public benchmark data-sets (SMLM2016) \cite{smlm2016}, they perform even worse than single-emitter algorithms for 3D data. These limitations imply that current fitting-based approaches to SMLM can not be applied to experiments with high emitter densities, which would be critical for the investigation of living or moving structures.
Furthermore, most previous algorithms base their predictions on a single observed image. Thus, they ignore potentially useful information in the sequence of imaging frames which can enable detecting and separating fluorophores in crowded high density data by taking into account their temporal dynamics.  Nevertheless, attempts at using information from multiple images during inference are rare \cite{cox2012bayesian, sun2016scalable}, and have not yet yielded state-of-the-art performance.

Deep learning methods have revolutionized computer vision, and biological image analysis is no different \cite{alexnet, review_royer, opportunities, pushing}. 
Many of these advances are the result of the supervised training of deep neural networks using large training datasets of pairs of example input images and desired output predictions.  While the analysis of SMLM data is not a standard supervised learning problem, ground truth localization data for training a deep network can be generated by simulating the imaging of fluorophores. The two first applications of deep learning to SMLM, DeepSTORM3D \cite{deepstorm} and DeepLoco \cite{deeploco} took this approach and used simulated synthetic SMLM to train deep networks to localize single molecules, an approach we call ``simulator learning'' (SL) \cite{le2017modelbased} in this paper. These two deep learning methods differ in the output representation used by the networks. DeepSTORM3D directly predicts a high resolution 3D volume which has the advantage of simplicity but the disadvantage that increasing the resolution of the predictions requires increasing the computation. In contrast, DeepLoco predicts continuous localizations for a fixed number of particles, and has the advantage that its computational complexity scales with the maximum number of possible particles, rather than size of the volume predicted. While both approaches produce detection uncertainty, neither predicts localization uncertainty.

We present a new method for fast, efficient, and accurate single-molecule localization based on a new deep neural network architecture we call DECODE (DEep COntext DEpendent) which achieves state of the art performance. DECODE also uses simulator learning, but is based on three main innovations:  First, we introduce a novel network architecture which uses temporal context for inferring fluorophore locations.  This single DECODE network is trained to produce accurate predictions at both low and high densities, alleviating the need for analysis methods which deal with these `single emitter' and `multi emitter' cases separately. Second, we phrase localization as a Bayesian inference problem, and provide a novel cost-function which makes it possible for the DECODE network to also predict uncertainty-estimates for each localized fluorophore. These uncertainty-estimators can, for example, be used for post-processing algorithms. Third, simulator learning depends on the faithfulness of the generative model, and might show reduced performance when there is a mismatch between the simulated and experimental data. We provide an alternative training approach, Combined Learning (CL) which combines Simulator Learning and Variational Auto Encoder learning (AEL)  \cite{vae_kingma, vae_rezende}, and evaluate its performance on simulated data.

We apply DECODE to data-sets from the public SMLM 2016 challenge \cite{smlm2016}, and show that it outperforms all existing methods which have been evaluated on this challenge so far, on $12$ out of $12$ data-sets for which DECODE is applicable \cite{smlm2016}. Our DECODE method leads to a improvement in performance which is $7\times$ as big as the improvement of the second best algorithm over the third best algorithm. Performance benefits are particular pronounced on high-density data-sets, on where the advantage from using DECODE increases to $10\times$ over the next best method . We also apply DECODE to four datasets where the same sample of labeled Tubulin-A647 protein was imaged with different densities of fluorophore activation, and demonstrate that we achieve high quality reconstructions with 10x less imaging time by accurately localizing fluorophores at high densities. To demonstrate the flexibility of our approach, we adapted it to reconstructing a large 3D volume of an entire COS7 cell with intracellular membranes densely labeled using PAINT, and imaged by lattice light sheet microscopy, imaged over several days. Our method significantly improved the reconstructions, but also enabled high quality reconstructions with only a fraction of the imaging time. Finally, we explore the performance benefits brought about by the use of local context, and the different training approaches.


\section{Results}

\subsection*{DECODE network for simultaneous detection and localization of fluorophores}

We designed and trained a deep neural network to simultaneously detect and localize fluorophores in SMLM measurements. The input to the deep network is a sequence of image frames containing sparsely activated fluorophores, and the desired outputs are locations of an unknown number of active fluorophores in each frame. 

\paragraph{Deep network design for predicting detection, localization, and uncertainty using spatial and temporal context} 
Previous deep learning approaches to SMLM have processed each frame independently, using one of two approaches, which we combine in our work. DeepSTORM3D \cite{deepstorm} produces predictions on a super-resolved 3D voxel grid. For each super-resolution voxel, a detection probability is predicted for the presence of a fluorophore. DeepLoco \cite{deeploco} combines classification and regression by predicting a fixed sized $256\times 4$ matrix for each imaging frame, with each row representing the presence or absence of a fluorophore, followed by the 3D vector containing the $x$, $y$ and $z$ coordinate of the molecule. This has the advantage that the computational complexity of the output scales only with the maximum number of possible active fluorophores in any frame, but not the volume of imaged field of view. However, it requires the network to learn a highly non-local and non-linear transformation from images into 3D coordinates in an undetermined ordering.

The DECODE network architecture is a hybrid of these two approaches: for each image frame it predicts eight channels for each imaged pixel (Fig. \ref{fig:concept}c). The first two channels indicate the detection probability of a fluorophore near that pixel in an imaging frame $p$ and its brightness $\alpha$. The next three channels describe the continuous valued localization of the fluorophore with respect to the center of the pixel, $\Delta x$, $\Delta y$, $\Delta z$. This hybrid approach allows DECODE to scale only with the number of imaged pixels (not super-resolution-pixels), and avoids a highly nonlinear and non-local mapping of pixels to coordinates.

DECODE is the first approach to provide fully probabilistic prediction of both fluorophore detection and localization. In addition to the first 5 output channels, three further  channels estimate the uncertainty of the localization along each coordinate given by $\sigma_x$, $\sigma_y$, $\sigma_z$. A final channel represents the uncertainty in the DECODE prediction of the brightness $\sigma_\alpha$. Thus DECODE directly predicts and independently represents uncertainty about detection, localization, and particle brightness.

\paragraph{Using temporal context}
We introduce a new mechanism to integrate information across frames, and show that it leads to improved detection and localization. The temporal dynamics of the fluorophores are such that a fluorophore can be active across multiple adjacent frames, inducing correlations which are local in time. We designed the DECODE network architecture (Fig. \ref{fig:concept}c) to infer the hidden states from three consecutive images and then use the combined information for the final localization. Using context has a substantial positive impact on performance.
%

\paragraph{Training the DECODE network using simulator learning} 
We want to train the parameters of the DECODE network to simultaneously detect and localize fluorophore particles from images of sparsely activated fluorophores. As ground truth particle localizations are not easily available for real data, we can not directly use supervised learning. However, the forward image formation model of how a given set of fluorophores gives rise to the detected image is well understood. To detect and localize active fluorophores, we need to train a neural network to invert this forward model. We investigate two different approaches for training the DECODE network to do this: The first method simulates data from our forward model and uses the simulated data to train the deep network using supervised learning \cite{le2017modelbased,cranmer2020frontier,gonccalves2020training}. We call this method ``simulator learning'' (SL, Fig. \ref{fig:sim_exps}a). The advantage of the method is its simplicity.  However, its accuracy depends crucially on the quality of the simulation and how well it matches the dataset being analyzed. We will describe a second method (called ``auto encoder learning'', AEL, Fig. \ref{fig:sim_exps}a) below. 

Simulator learning has been used by previous deep learning approaches to SMLM \cite{deeploco,deepstorm}. Since the physics describing how the camera image is generated by the imaging of a biological sample is well understood \cite{psf_accuracy}, we can use a simulation of biological samples consisting of point source emitters representing active fluorophores distributed randomly across a small image patch. We then model the forward generative process of image formation as follows: We simulate the noise- and background-free image of the fluorophores by convolving the point emitters with a model of the point spread function. A random homogeneous background intensity is added to generate a mean intensity image, and finally the noisy measured camera image is then simulated by sampling from a gamma distribution. The density, brightness, activation and inactivation times of the simulated fluorophores, and the background intensity values are chosen randomly to generate a large diversity of simulated images.

We developed a specialized loss function for our representation of the final localizations by the discrete pixel positions and the in-pixel offset variables.
 We interpret the binary values $p$ as the probability that an activation exists in that pixel while the outputs $\alpha,\Delta x,\Delta y,\Delta z,\sigma_\alpha,\sigma_x, \sigma_y,\sigma_z$ parametrize Gaussians which are components of a Gaussian mixture model (GMM) which describes the spatial distribution of emitter activations.
We then maximize the likelihood of the simulated continuous ground truth positions under this GMM. This allows us to optimize all the output variables jointly and to obtain uncertainty estimates which can be used to filter out localizations or to convolve the localizations with a Gaussian parametrized by the uncertainty for improved rendering \cite{visualization}. 
To determine the correctness of our uncertainty estimates we compared them to parametric estimates of the Cramér–Rao bound obtained with the equation from \cite{crlb}. Such estimates generally only take the brightness, the background and the PSF shape into account but not other important factors that increase uncertainty like other close-by PSFs or inhomogeneous background. We observe that under optimal conditions, with a single emitter per frame and a Gaussian PSF, our uncertainty estimates agree well with the parametric estimates. For denser data our method generally produces higher uncertainties, except when temporal context is used (see Fig. \ref{fig:crlb}). 


\begin{figure}
\includegraphics[width=1.0\textwidth]{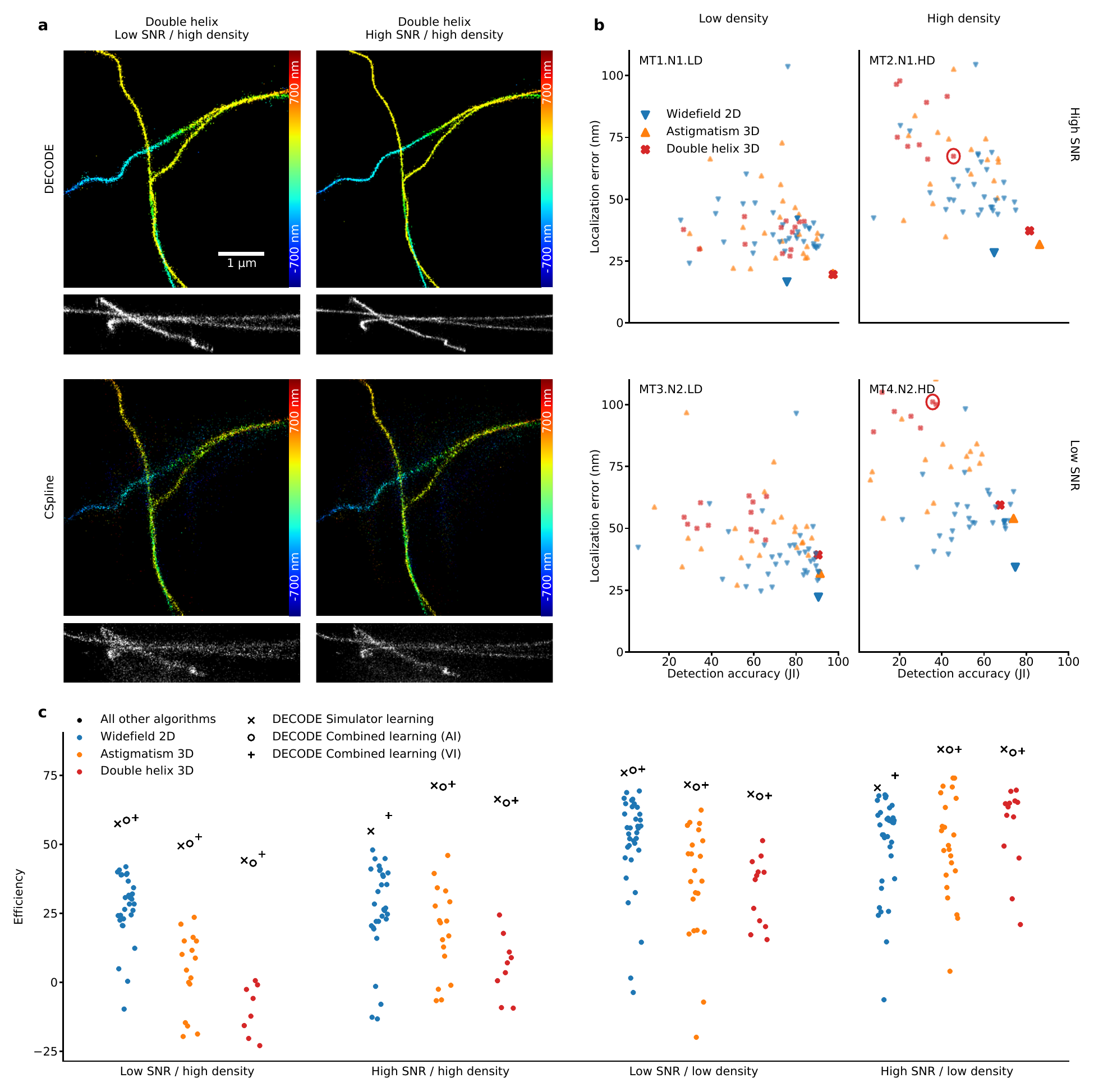}
\caption{
{\bf Performance comparison on the SMLM 2016 challenge.} {\bf a)} Reconstructions by DECODE and the CSpline algorithm on high density double helix challenge data.
 Upper panelsc olor coded $x$-$y$ view, lower panels $x$-$z$ cross section. {\bf b)} Performance evaluation on the twelve test datasets with low/high density, low/high SNR and different modalities using the Jaccard index (higher is better) and lateral localization error (lower is better) as metrics. Each marker indicates a benchmarked algorithm, large solid markers indicate DECODE, red circles indicate CSpline results for the conditions in Panel a.
 {\bf c)} Performance scores for three DECODE variants (see section on combined learning) across all twelve data-sets, quantified with efficiency scores. Colored dots indicate performance numbers for other methods taken from challenge website.} 
\label{fig:results_challenge}
\end{figure}

\subsection*{Quantitative evaluation on simulated datasets from the SMLM challenge show DECODE outperforming all algorithms across a variety of conditions}


The 2016 SMLM challenge \footnote{http://bigwww.epfl.ch/smlm/challenge2016/index.html?p=datasets} is the second generation comprehensive benchmark evaluation developed for the objective, quantitative evaluations of the plethora of available localization algorithms \cite{smlm2014,smlm2016}. The benchmark offers synthetic datasets for training and evaluation that were created to emulate various experimental conditions. 
%
A direct comparison of DECODE with other contenders (Fig. \ref{fig:results_challenge}) in the SMLM 2016 challenge shows DECODE outperforming other approaches across datasets \footnote{Note that currently not all of our results are displayed in the plots on website, but all results can be downloaded from http://bigwww.epfl.ch/smlm/challenge2016/leaderboard.csv. Our results uploaded January 16 2020, results are current as of July 06 2020.}.
DECODE outperforms all 39 currently ranked algorithms on 12 out of 12 datasets, and often by a substantial margin. The datasets include high (N1) and low (N2) signal to noise ratios (SNR), with low (LD) or high (HD) emitter densities, with 2D, Astigmatism (AS) and Double Helix (DH) point spread function based imaging modalities Fig. \ref{fig:results_challenge}. We quantified performance using RMSE lateral or volume localization error, as applicable for 2D and 3D data respectively, and the Jaccard index $JI$ which measures single molecule detection accuracy. The SMLM 2016 benchmark also reports a single score which combines particle localization and detection accuracy into a measure called efficiency.

DECODE achieves an average efficiency score of 66.61 out of the best possible score of 100 (achievable only by a hypothetical algorithm that accurately detects 100\% particles with 0 \si{nm} localization error). This is compared to an average score $48.3$, and $45.6$ for all non-DECODE second and third place algorithms respectively. The improvement in performance by using DECODE is substantial, leading to 7 $\times$ the accuracy improvement gained by using the second best algorithm over the third best algorithm.
The difference is particularly large under difficult imaging conditions, when high emitter densities and low SNR can conspire to make detection and localization challenging, particularly so for the double helix point spread function. For example, in the Low SNR/high density/Double Helix condition, DECODE achieves an efficiency score of 44.23, whereas no other algorithms achieves a non-negative efficiency score. DECODE achieves an average efficiency of 57.29 on the six challenging high density datasets, while the average second best and third best algorithms achieve only scores of 30.76 and 27.16 respectively. This represents an average 10 $\times$ improvement relative to the improvement by the second best algorithm over the third best algorithm on these challenging datasets.

DECODE is the best algorithm on all 12 datasets, across a variety of imaging modalities, SNR and density conditions. In contrast no other algorithm previously achieved such universal superiority, instead specializing on a limited range of imaging conditions.
Qualitatively, DECODE improves super-resolution reconstructions by improving both the detection and the localization of single molecules. An example of this can be seen in Fig. \ref{fig:results_challenge}a, where we compare the reconstructions obtained with DECODE to the multi-emitter fitting approach CSpline \cite{Spliner} on two 3D double-helix datasets with high fluorophore densities \footnote{We used settings provided by the authors: https://github.com/ZhuangLab/storm-analysis}. DECODE detects more fluorophores, and localizes them more accurately than CSpline for this dataset.


\begin{figure}
\includegraphics[width=1.0\textwidth, trim={.cm 0.3cm 0.0cm 0.0cm},clip]{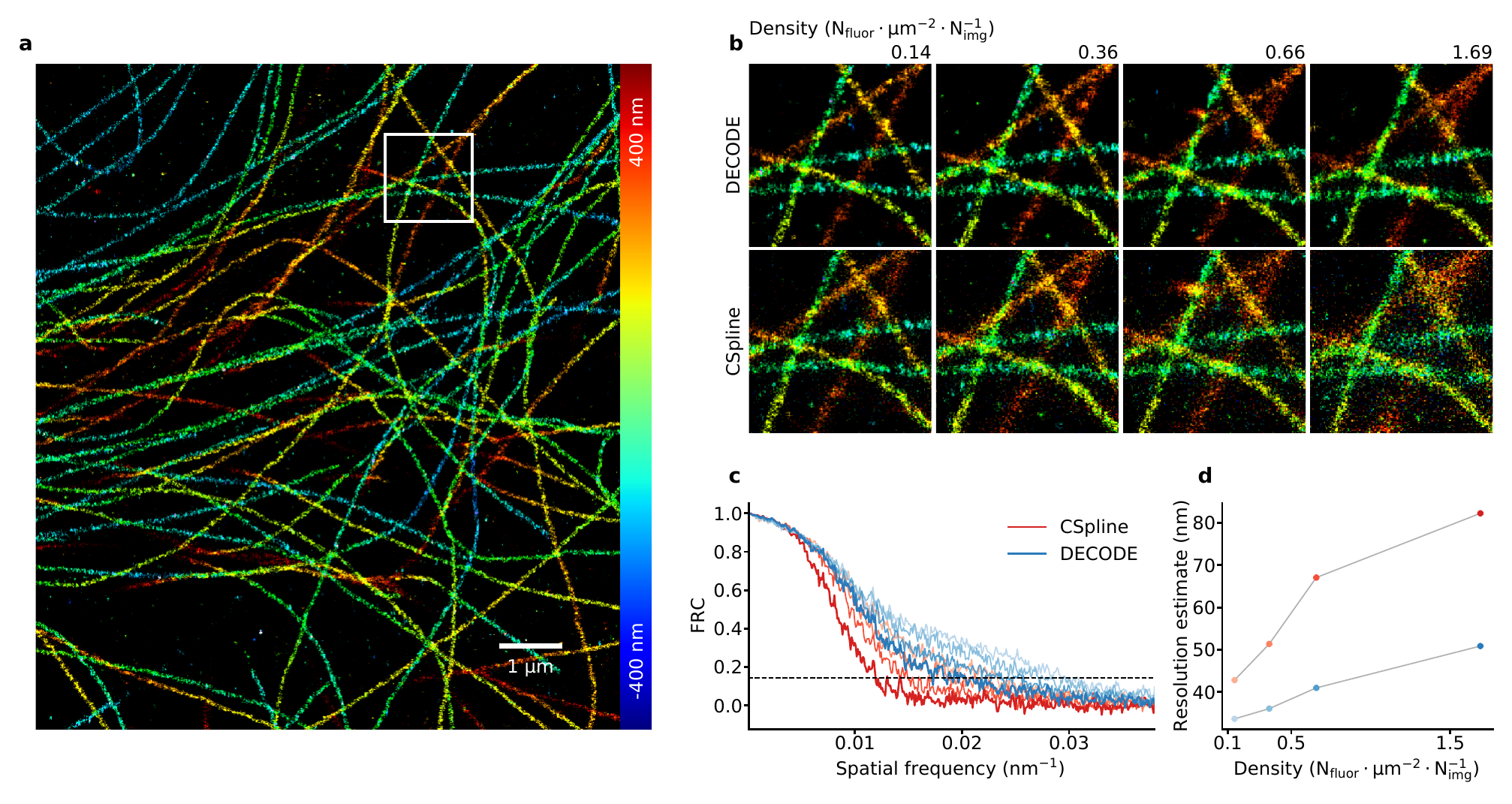}
\caption{
{\bf DECODE produces superior reconstructions at higher densities with fewer frames of data.} 
{\bf a)} Reconstruction of one section of the low density dataset with DECODE. 
 {\bf b)} Magnified reconstructions (corresponding to boxed section in panel a), for different densities obtained with DECODE and CSpline. 
 {\bf c)} Resolution estimates obtained using the Fourier Ring Correlation and 0.143 criterion across densities for both methods.
}
\label{fig:results_ries}
\end{figure}

\subsection*{DECODE enables accurate reconstructions with shorter imaging times at high emitter densities}

In SMLM, there is a trade-off between the imaging time and the activated fluorophore density. Sparsely activating fluorophores leads to the best localization accuracy, but requires long imaging times in order to localize sufficient numbers of particles to reconstruct the sample faithfully. By enabling accurate particle localizations at higher densities, DECODE can yield accurate super-resolution reconstructions with significantly shorter imaging times. We demonstrate this by imaging and reconstructing the same sample of labeled microtubules at four different emitter densities using dSTORM (direct stochastic optical reconstruction microscopy) \cite{dstorm}.
For the first dataset the experimental conditions roughly correspond to high SNR and low density settings modelled in the challenge, with an average upper limit of emitter density of 0.14 fluorophores $\mathrm{\mu m}^{-2}$ per image \footnote[1]{Measured by dividing the DECODE predictions into 1 $/\mu m^{-2}$ bins and calculating the 99 percentile of densities.}.
The other three datasets consisted of 2.5 $\times$, 4.5 $\times$ and 12 $\times$ fewer imaging frames, while the total number of active fluorophores is roughly constant across datasets. 

We trained and applied one common DECODE model to all four datasets (Fig. \ref{fig:results_ries} a: reconstruction on subsection of low density dataset). We compared DECODE reconstructions with those of CSpline across all four data sets (Fig. \ref{fig:results_ries} b) to investigate how the quality of reconstruction deteriorates for denser datasets. Similarly to the simulated challenge datasets we observe a sharper image and less spurious localizations for the original low density dataset. As the density increases, DECODE consistently yields reconstructions with similar accuracy, while the reconstructions produced by CSpline degrade for high densities.

To quantify the reconstruction performance across the different conditions, we calculated the resolution of the reconstructed image using Fourier Ring Correlation (FRC) \cite{frc}. The FRC estimates resolution by measuring the correlation of two different reconstructions of the same image across spatial frequencies. We split the localizations ordered in time into blocks of 10000 and created two different reconstructions of the same sample coming from even and odd blocks. We then used the spatial frequency at which the correlation drops below a threshold value of 0.143 \cite{frc_crit} to estimate the resolution of the reconstruction. DECODE consistently improves resolution by 10 \si{nm} - 25 \si{nm} over CSpline across all imaging densities (Fig. \ref{fig:results_ries}c and d), and requires $10 \times$ fewer imaging frames for the same quality of reconstruction. 


\subsection*{DECODE enables high fidelity reconstructions of 3D lattice light sheet PAINT imaging with reduced imaging time}

To illustrate the general applicability of DECODE, we applied it to 3D lattice light sheet (LLS) microscopy combined with the PAINT (point accumulation for imaging of nanoscale topography labeling) technique \cite{lls-chen, paint}. In PAINT microscopy, the fluorophore labeling a sample stochastically binds and unbinds from the sample, providing dense labeling. In lattice light sheet microscopy, thick volumes can be imaged at high resolution by scanning a thin ($1.1 \si{um}$)  light sheet, with axial localization within the sheet enabled by astigmatism. We reconstructed a previously reported dataset of a chemically fixed COS-7 cell with intracellular membranes preferentially labeled by azepanyl-rhodamine (AzepRh) \cite{lls-cos7} consisting of $147,500$ 3D volumes comprising more than 20 million 2D images acquired in $270\si{nm}$ steps.

For this dataset,  one complete scan of the volume involved moving the probe 141 times by $500 \si{nm}$. The detection axis was oblique to the coverslip, resulting in emitters that are active in successive frames, and thus appear to move in the x- and z-direction by a fixed distance. We adjusted our algorithm to account for this movement, so that we could still employ local context. Furthermore, we used a modified noise model, as the images were made with a sCMOS camera (see methods for details).

\begin{figure}
\includegraphics[width=1.0\textwidth, trim={.cm 0.3cm 0.0cm 0.0cm},clip]{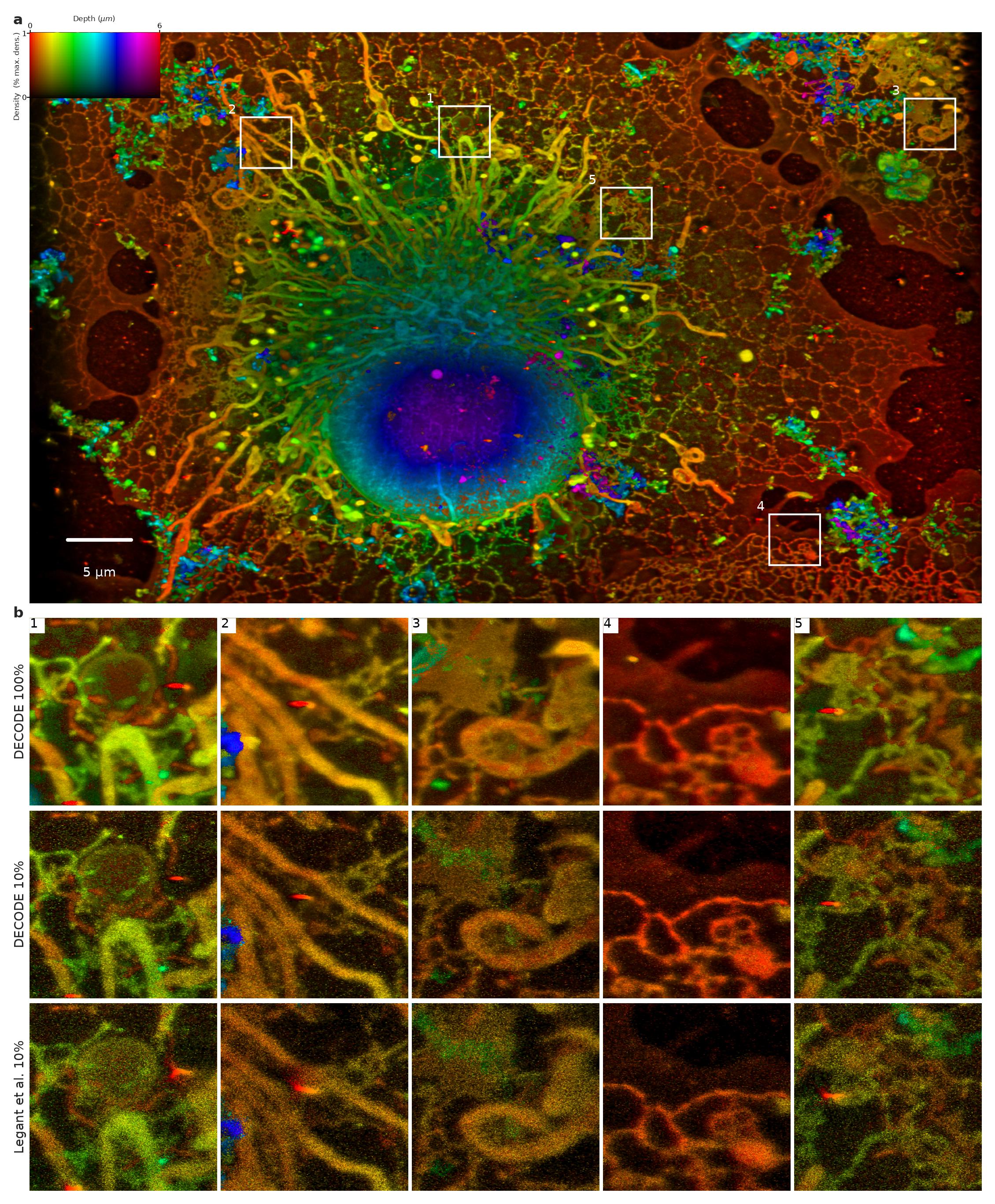}
\caption{
{\bf DECODE resolves structural details with only 10 \% of frames in LLS-PAINT dataset.} 
{\bf a)} COS-7 cell imaged with LLS-PAINT microscopy. 
Viewing angle lies perpendicular to the specimen. 
 {\bf b)} Magnified reconstructions of boxed sections in panel a. First row: DECODE renderings on all 147,500 recorded volumes.
 Below: Renderings of DECODE and those provided by Legant et al. \cite{lls-cos7} from 10 \% of the available volumes. {\bf 1:} DECODE resolves the hollow structure of endosomes more clearly. {\bf 2:} DECODE avoids distortions of structures around fiducials. {\bf 3:} Mitochondrial substructures (like the cristae in yellow) are better resolved. {\bf 4:} DECODE can help to distinguish whether ER tubules are continuous or broken. 
}
\label{fig:lls}
\end{figure}

We compare our reconstructions to the original reconstructions described in \cite{lls-cos7} which used a  custom-made iterative MLE fitter with a parametric PSF model \cite{lls-cos7}. DECODE detects $1.25$ billion particles, compared to $400$ million particles detected by the original algorithm.
While LLS-PAINT microscopy yields high resolution reconstructions over large 3D volumes, its usability is limited by the long imaging times required to localize a sufficient numbers of particles for reconstruction. For example, the dataset we analyzed was obtained in over 3 days of imaging time. We show that DECODE provides sharper images using a smaller number of frames, and could thus be used to obtain the same quality of reconstructions using only a fraction of recorded frames which is confirmed by FRC resolution estimates (Fig. \ref{fig:lls}, \ref{fig:lls_supp}).  
We note that one challenge of long imaging times is that performance can be limited by nonlinear swelling of the sample over the time course of the imaging, which can only be partially corrected by non-rigid registration. Thus, reducing imaging time by improved reconstruction algorithms could also lead to better reconstructions with fewer artifacts. %


\subsection*{Combining simulator learning with auto encoder learning to learn to simulate better}
The effectiveness of simulator learning depends  on the availability of an accurate forward generative model at training time, as deviations of the true forward model from the simulated forward model can degrade performance. This problem can be solved by simultaneously estimating the parameters of the true forward model, and training the DECODE network using the real measurements, rather than a fictitious simulation. This is possible using the recently developed framework of variational autoencoders (VAEs) \cite{vae_kingma, vae_rezende}. In the VAE framework, the stochastic forward generative model and the DECODE network are stacked to form a stochastic autoencoder. This autoencoder is then used to simultaneously optimize the parameters of the deep network and the forward model, with the goal of achieving image-reconstructions which are similar to the original measurement.
 
Formally, this can be achieved by maximizing a so-called `evidence-lower bound' via stochastic gradient optimization (see methods for details). VAEs have, e.g.,  previously been used on the related problem of inferring action potentials from calcium imaging data \cite{speiser2017}. A drawback of VAE-based approaches are that gradients for training the DECODE network need to be approximated using Monte Carlo sampling, which can make optimization more challenging.

Simulator learning and autoencoder learning are two sides of the same autoencoder coin, as can seen by comparing Fig 4a to Fig 4b: In simulator learning, a known PSF is used to transform (encode) simulated emitter locations into a microscope image, and the DECODE network is used to recover (decode) the original simulated emitter locations. The network is optimized to minimze the discrepancy-measure computed by comparing simulated and inferred localizations.

In what we call autoencoder learning (due to the relationship to VAEs) the image measured by the microscope is stochastically transformed (encoded) by the DECODE network into predicted emitter locations, and these predicted emitters are then transformed (decoded) by the estimated PSF back into a reconstruction of the measured image. The objective function used to train the DECODE network is the difference between the measured and reconstructed images. Because AEL relies on autoencoding the real measured data, it enables learning of both the encoder (DECODE network) and the decoder (PSF). In contrast, SL only allows training of the decoder (DECODE network). Empirically, we found that combining SL and AEL in a manner we call combined learning (CL) often leads to the best performance. 

\begin{figure}[H]
\includegraphics[width=1.0\textwidth,trim={0cm 5.0cm 0.0cm 0.0cm},clip]{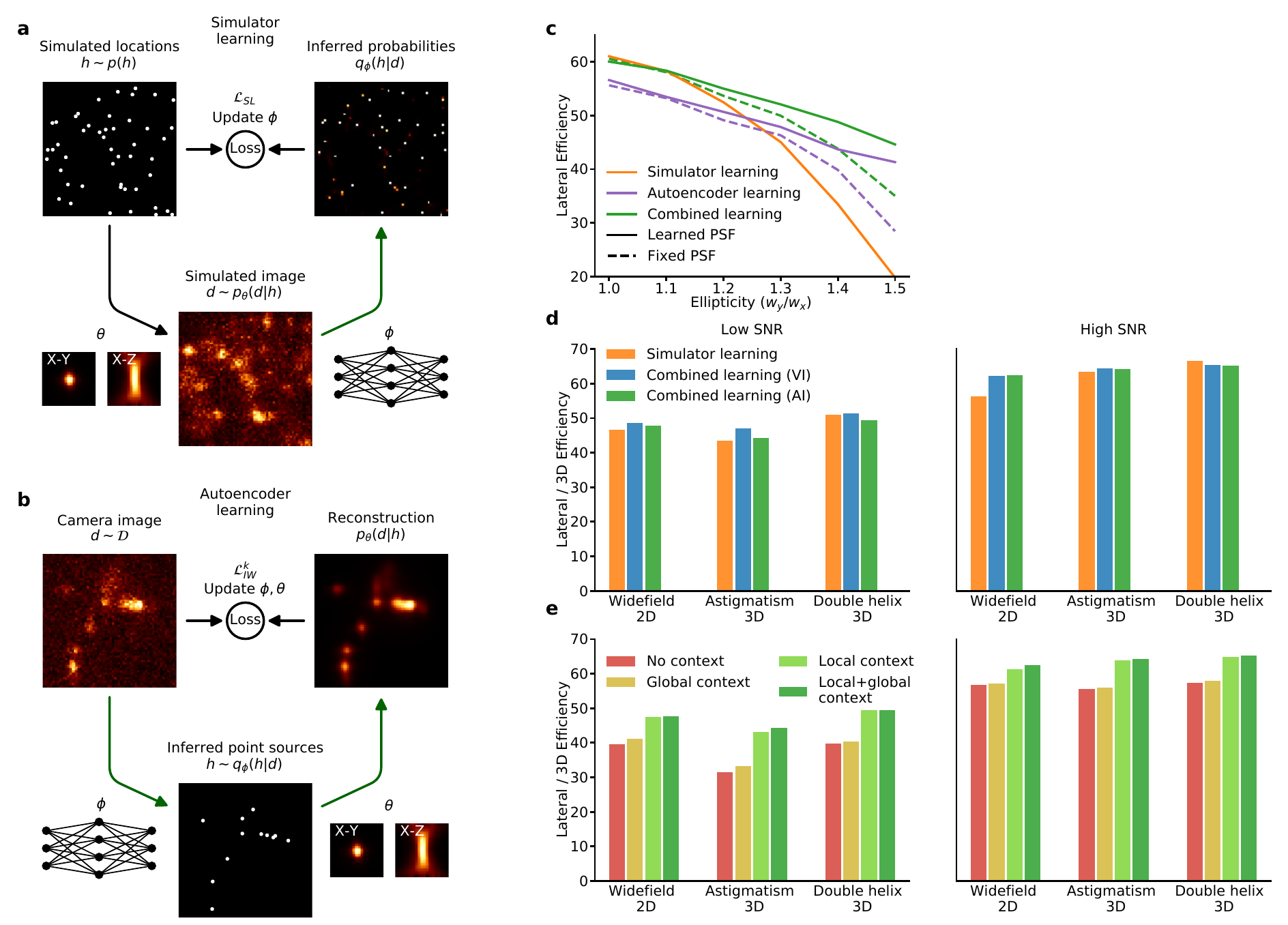}
\caption{{\bf DECODE performance for training approaches and settings.
a)} Simulator learning (SL). Synthetic images are constructed by the simulated imaging of randomly located fluorophore point sources
using a generative model, and a network is trained to detect and localize the fluorophores using supervised learning. 
{\bf b)} Auto-encoder learning (AEL). A neural network used to infer putative locations from a measured camera image, and subsequently the generative model is used to reconstruct the original camera image. Both the parameters of the generative model and of the DECODE network are optimized. Loss is computed between measured end reconstructed images.
{\bf c)} Performance of different training methods for different degrees of PSF mismatch. Models using a circular PSF are fit to 5 datasets simulated from PSFs with varying ellipticity. PSF parameters for AEL / AEL+SL learning could be either fixed (solid line) or learned (dashed line).
{\bf d)} Performance of DECODE trained with different methods on the 6 high density challenge test datasets.
{\bf e)} DECODE evaluated on the 6 high density challenge training datasets. Models were trained using combined learning either without context, local, global and
both forms of context. 
\label{fig:sim_exps}}
\end{figure}

To highlight the difference between simulator and autoencoder learning we show how the two approaches behave for different degrees of mismatch in the assumed point spread function. 
To simulate PSF mismatch we generated datasets using a 2D elliptical Gaussian PSF with increasing ellipticity and then trained DECODE models that use a non-elliptical circular Gaussian (Left panel Fig. \ref{fig:sim_exps}e, solid lines show performance of models with fixed generative parameters). For an ellipticity of zero the models have access to the true underlying generative model. In this case SL training sets an upper bound to the achievable performance with a given network as it is able to generate an infinite amount of labelled data with the correct simulation parameters, and so outperforms AEL. 
However, pure simulator learning is brittle and more sensitive to parameter mismatch as the DECODE network never `sees' elliptical PSFs during training. Autoencoder learning can  still try to infer the correct positions as placing the circular PSF into the middle of the elliptic one achieves the best reconstruction. Alternating between the two methods retains the advantages of both: Performance is virtually the same as simulator learning when there is no mismatch and performance degrades more gracefully when the mismatch is increased. 
When we add training of the generative model parameters (dashed lines), our PSF model learns to account for some of the PSF-mismatch, which further improves performance. 

To test these findings in a more realistic setting, we trained DECODE methods using simulator and combined learning and submitted them to the SMLM 2016 challenge (see Fig. \ref{fig:results_challenge}). We also evaluated them on the training datasets to more precisely evaluate differences in performance (Fig. \ref{fig:sim_exps}d). For astigmatism and double helix data we fit the PSF model to the provided bead-data. For 2D we instead used a more heuristic estimate, choosing sigma values and a z-dependent scaling that covers the PSF observed in the dataset.
Furthermore we used combined learning in two different settings: one for variational inference (VI) where autoencoder learning is performed on the same dataset on which performance is evaluated and one for amortized inference (AI) where training and testing takes place on different datasets.
These two models where trained using both local and global temporal context, while for simulator learning we only used local context.
Overall performance of the three submissions is very similar, which is to be expected given that we are able to approximate the generative model of the data very accurately. 
Adding autoencoder learning is especially helpful for 2D data where we did not fit the PSF model to beads-data and instead adjusted it during training of the model. As can be seen in Fig. \ref{fig:wmap_plots} the algorithm learned to add diffraction rings to the PSF model without any access to bead-data. 
Additionally we observe the combined learning (VI) performs better on the high density / low SNR datasets. As described below, this can be attributed to the fact that global context is especially helpful in these difficult conditions.


\subsection*{Local temporal context is more informative than global context}
\label{sec:features}

Imaging frames from SMLM contain correlations in time across short and long time-scales. Once activated, fluorophores are usually active for more than one imaging frame and are therefore visible in multiple consecutive frames at the same position, leading to short time-scale correlations in the images. The spatial distribution of fluorophores in a sample is non-uniform and is concentrated around the biological structures labeled by the fluorophores. This spatial distribution of fluorophores leads to temporally global correlations. Our DECODE network is designed to exploit both kinds of temporal correlations via local and global context windows across imaging frames to improve the detection and localization of single molecules. We studied the contributions of local and global context to the performance of the DECODE network.

We trained DECODE models using CL on the 6 high density datasets of the challenge, using either no context, local context, global context or both forms of context.
Local and global context have different effects on the performance (Fig. \ref{fig:sim_exps}e). While we find that both local and global context individually improve performance, local context is generally more helpful than global context. Together, global and local context give the best possible performance but at a minor improvement over local context alone.

It should be noted that a straight forward way of using local context called ``grouping'' is commonly used to improve localizations as a post-processing step in SMLM \cite{grouping}. Localizations occurring in consecutive images that are closer to each other than a fixed threshold are assumed to belong to the same emitter and their localization is averaged, potentially weighted by the uncertainty of each localization. 
We applied grouping to the DECODE models trained with and without local context as well as CSpline (Fig. \ref{fig:grouping}) . We observe that grouping is almost as effective as our method in reducing the localization error for the easiest condition (high SNR / low density) but performs worse for any of the more difficult datasets. Furthermore, our method for using context also improves detection accuracy while grouping only influences the localization error. Lastly, using grouping on top of DECODE with local context results in a small improvement which would further increase  performance on the challenge if one used it. 



\section{Discussion}

We here described DECODE, a new deep-learning based method for single molecule localization for reconstructing super-resolution images. DECODE differs from existing localization algorithms by simultaneously performing detection and localization of particles. DECODE yields substantial improvements in performance over previously evaluated algorithms in a publicly available benchmark challenge: It achieves best performance in every condition, and often improves prediction performance by a large margin. When applied to high density dSTORM imaging of microtubules, and LLS-PAINT imaging of whole cells, it leads to reconstructions which have markedly improved resolution due to substantial improvements in particle detection. The performance benefits of DECDOE are especially pronounced in high-density imaging conditions, thereby opening up new opportunities for faster imaging of fixed samples, and even live imaging. 

DECODE leverages the flexibility of deep learning -- for its predictions, the network can benefit both from temporal context (e.g. from fluorophores being active across multiple imaging frames), as well as spatial context (e.g. from clustering of fluorophores in space). DECODE can be used in a very flexible and general manner and can easily be applied to arbitrary PSFs and noise models -- in this paper, we applied it to 4 different imaging modalities ranging from engineering point spread functions to 3D lattice light sheet microscopy.

The DECODE network is trained to produce probabilistic point process predictions -- it predicts both the probability of detection and the uncertainty of localization for each detected particle. We showed that the localization uncertainties predicted by our network are superior to conventionally used CRLB uncertainties and are particularly useful for filtering particles to produce high resolution reconstructions.

We presented and evaluated two ways to train the DECODE network, using simulator learning and autoencoder learning. Simulator learning allows for fast an easy training of a DECODE network when the optical properties of the microscope are precisely known. Autoencoder learning uniquely enables the \emph{in situ} estimation and refinement of imaging parameters such as the empirical point spread function and noise model directly from the experimental measurements, in principle enabling the tracking of drift in the optical system over the course of long imaging experiments, or the estimation local point spread functions across large fields of view in the presence of sample dependent optical aberrations.

One weakness of DECODE is that it currently requires the training of a new neural network whenever the optical properties of the microscope change. This training can currently take over 10 hours on a single GPU. However, it may be possible to train a single network to predict robustly across minor variations in the point spread function or noise distribution, i.e. to \emph{amortize} simulator-learning across setups \cite{speiser2017,gonccalves2020training}. This can enable real-time reconstruction on a single GPU without the need for re-training, even under the most challenging conditions, since the computational complexity of the predictions depend only on the size of the image and not the number of particles in each imaging frame.


\subsubsection*{Acknowledgments}

This work was supported by the German Research Foundation (DFG) through SFB 1089 and  Germany’s Excellence Strategy – EXC-Number 2064/1 – Project number 390727645, the German Federal Ministry of Education and Research (BMBF, project `ADMIMEM', FKZ 01IS18052 A-D), the Howard Hughes Medical Institute and the  European Research Council (CoG-724489 to J.R.). We thank Daniel Sage for useful discussions,  David Greenberg and Poornima Ramesh for comments on the manuscript, and Eric Betzig and Jennifer Lippincott Schwartz for kindly sharing data with us. 

\section{Methods}

\subsubsection*{Software availability}
All methods were implemented in Python and PyTorch \cite{pytorch}. Code and hyper-parameter settings for the challenge results are available at \url{https://github.com/mackelab/DECODE}/

\subsubsection*{Approximate Bayesian inference with DECODE}

DECODE is a Bayesian inference method which requires a formal probabilistic description of the entire SMLM measurement process. This description amounts to a stochastic simulator (known as a generative model) which can generate synthetic SMLM data, but which can also be optimized to fit the data. In our framework, at each time point, each active fluorophore $i$ has a location in 3D at $x_i$, $y_i$, $z_i$, and a brightness $\alpha_i$. The set of $N_t$ single molecule locations and their brightness at each time point are the unknown hidden causes or latent variables $h = \{\{x_i\}_t,\{y_i\}_t,\{z_i\}_t,\{\alpha_i\}_t\}$ that give rise to the noisy low resolution images which constitute the measured data $d = \{I_t\}$.
A complete description of the generative model in Bayesian framework includes the prior distribution $p(h)$ which describes the spatial distribution and temporal dynamics of fluorophores, and the likelihood distribution $p(d|h)$ which describes the stochastic process describing the distribution of images generated by the microscope for a given configuration of fluorophores. The likelihood is formalized in terms of the point spread function describing the transfer function of the microscope, and the measurement noise at the camera.

The DECODE network is trained to perform approximate Bayesian inference, using simulator learning and autoencoder learning to approximate the true posterior distribution $p(h|d)$ and predicts the hidden single molecule locations and brightness from the measured images.

\subsubsection*{Spatial distribution and temporal dynamics of fluorophores.}
We assume that on each imaging frame, a fluorophore can be activated in any given pixel with a constant probability of $p_{on}$. An active fluorophore on any given frame has a probability of $p_{off}$ of turning off in the next frame. The location of an active fluorophores within a pixel is drawn from a uniform distribution in $x$ and $y$. For $z$ we chose a Gaussian distribution with a mean centered at the focal plane and a variance chosen to cover the range of the point spread function of the microscope instead, as the bulk of the recorded structure is usually located around the focal plane. The brightness of an active fluorophore is sampled from a uniform distribution $U(0.1,1)$ times the maximum possible expected brightness of a single fluorophore. This describes the prior distribution over particle locations and brightness $p(h)$.

\subsubsection*{Point spread functions} 
\label{sec:psf_model}
We used the sum of a parametric function $PSF_{parametric}$  and a non-parametric interpolated pixel map $PSF_{pixmap}$ to model arbitrarily complex point spread functions $PSF(x,y,z)=PSF_{parametric}(x,y,z) + PSF_{pixmap}(x,y,z)$. In principle, the non-parametric interpolated pixel map is sufficient to represent any possible PSF within the support of the pixel map, we found that the parametric component helps with the learning when using AE training.

In this paper, we analyzed data imaged or simulated using three PSF models -- one for 2D localization, and two for 3D localization. For 3D localization, we used the astigmatism (AS, \cite{huang2008three}) and the double helix (DH, \cite{pavani2009three}) PSFs. The parametric components of these parametric PSFs as a function of the spatial coordinates $x$, $y$, and $z$ is given below, along with the parameters of the PSF after the semi-colon.

\begin{align}
\label{eq:psf}
PSF_{2D}(x,y,z; {a_1,a_2}, {b_1,b_2}) = & \sum_{n=1,2}(a_n \cdot e^{-b_n(x^2+y^2)/(1+|z|)^2})\\
PSF_{AS}(x,y,z;a,b_x,b_y,c) = & e^{-ax^2/((z-b_x)^2+c)}e^{-ay^2/((z-b_y)^2+c)} \\
PSF_{DH}(x,y,z;a,b,c,d) = & e^{-a(x-s_x(z))^2+a(y-s_y(z))^2} + e^{-a(x+s_x(z))^2+a(y+s_y(z))^2} \\
&s_x(z)= d\cdot\cos(b \cdot z + c) \quad s_y(z)=d\cdot\sin(b \cdot z + c).
\end{align}

$PSF_{2D}$ is a weighted sum of two circular Gaussians whose variance increases as a function of the distance of the point source away from the focal plane. $PSF_{AS}$ is described by an elliptical PSF whose eccentricity is a function of the defocus $z$ \cite{daostorm}. $PSF_{DH}$ is a double helical PSF modelled as two circular Gaussians that rotate around the fluorophore at a distance $d$ as a function of $z$.

$PSF_{pixmap}$ enables the approximation of arbitrarily complex PSFs and is a 3D image volume with the same pixel size as camera and with a z-spacing of 100 \si{nm}. This pixel map is interpolated by trilinear interpolation to evaluate the point spread function at any location within the support of the pixel map.

\subsubsection*{Imaging fluorophores and camera noise} 
The number of photons emitted from a fluorophore follows a Poisson distribution. To model the noisy imaging by an EMCCD camera, the distribution of photon counts can be convolved with a gamma distribution that models the electron-multiplying (EM) gain and with a Gaussian distribution that accounts for read-out noise. 
The resulting distribution can not be expressed analytically. This is irrelevant for simulator learning, however autoencoder learning requires a differentiable expression for the probability distribution of the measured image $I(x,y)$ as a function of the mean intensity image $\bar{I}(x,y)$. Therefore, we approximate the noise model using a single Gamma distribution with parameters that are given by the camera baseline $BL$, its electron-multiplying gain $EM$ and electron conversion factor $EC$. 
Given $N$ activated fluorophores, each located at $x_n,y_n,z_n$ and with a brightness of $\alpha_n$, and a constant background fluorescence of $\beta$, the intensity of a pixel $I(x,y)$ located at $x,y$ of the resulting imaging frame is  simulated as:
\begin{align}
\label{eq:emccd}
\bar{I}(x,y) =& \sum_{i=1}^N \alpha_i PSF(x-x_i,y-y_i,-z_i) + \beta \\
I(x,y) \sim& \textrm{Gamma}((\bar{I}(x,y) - BL)/\eta, \eta) + BL\\
&\eta = 2\cdot EM/EC.
\end{align}
This describes the likelihood function $p(d|h)$.


\subsubsection*{Fitting of 3D PSFs from bead stacks}
\label{sec:psf_fit}
\begin{figure}[h]
\includegraphics[width=1.0\textwidth]{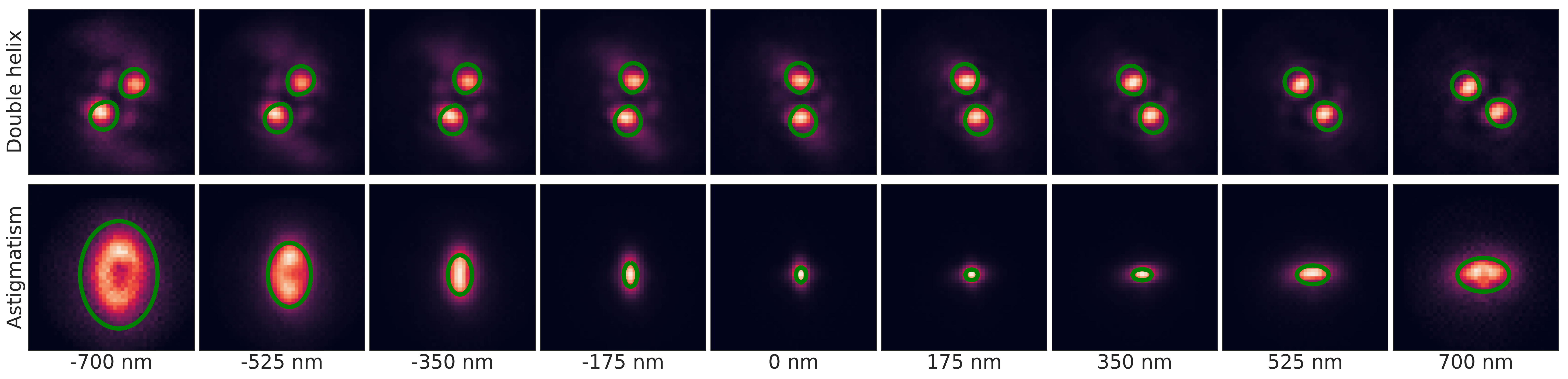}
\caption{
{\bf PSF model fits} PSFs were fitted on astigmatism and double helix challenge data. Contour plots show the underlying parametric model and highlight the contribution from the pixel maps $\delta_{xy}$.}
\label{fig:psf_plots}
\end{figure}

\begin{figure}[h]
\includegraphics[width=1.0\textwidth]{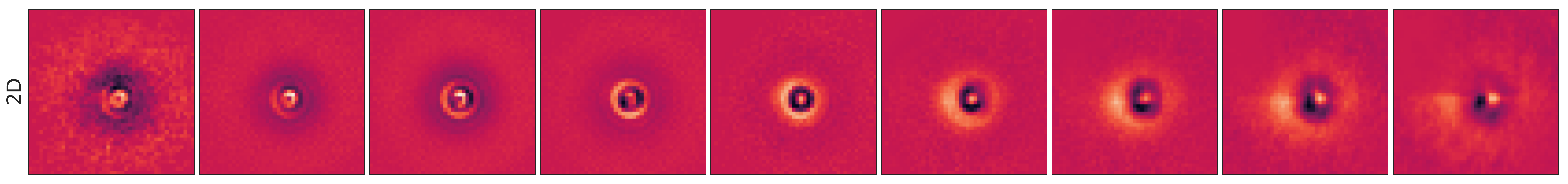}
\caption{
{\bf Learned pixel maps} $\delta_{xy}$ for different absolute z-offsets learned when using combined learning on the high SNR widefield 2D challenge dataset. The model was clearly able to identify inference rings without the use of calibration bead stacks.}
\label{fig:wmap_plots}
\end{figure}

\begin{figure}[H]
\includegraphics[width=1.0\textwidth,trim={0cm 0.0cm 0.0cm 0.0cm},clip]{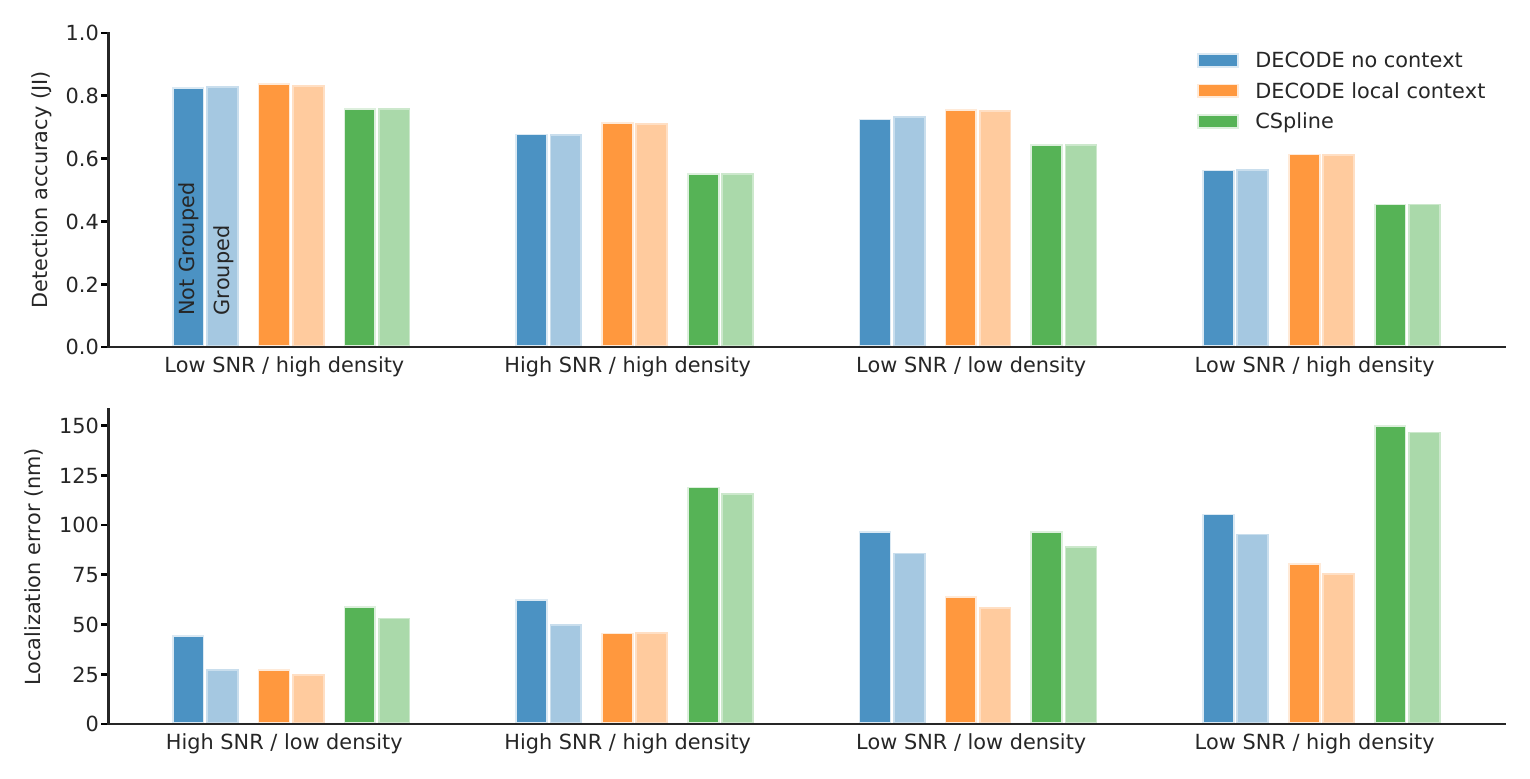}
\caption{{\bf Comparison of the impact of local context and grouping} 
Performance on the four astigmatism challenge datasets for DECODE models trained with and without local context, as well as CSpline. For each algorithm detection accuracy and RMSE are shown for raw and grouped predictions. Across all conditions DECODE with local context and without grouping outperforms DECODE without context as well as CSpline when using grouping. 
\label{fig:grouping}}
\end{figure}

For 3D inference, it is common to calibrate the PSF model on data with known axial offset as the exact relationship between its shape and the position cannot be estimated from unlabeled data. We estimated the AS and DH PSF's using calibration bead stacks, i.e. images of single fluorophores at different offsets with  high signal to noise ratios. We first obtain a rough estimate of the bead locations using a basic peak-finding routine. We then maximize the likelihood $p_{\theta}(d|h)$ by performing stochastic gradient descent on the exact $x-$ and $y-$ coordinates of each bead (which are constant across images), the shape parameters of the PSF model (\ref{eq:psf}) and the the pixel maps $\delta_{xy}$. This simple method achieves localization errors of less then 0.3 \si{nm} on the challenge calibration stacks where the ground truth locations are available.
During training of the DECODE network we generally keep the PSF model fixed (see example fits in Fig. \ref{fig:psf_plots}). For 2D datasets the PSF model can be learned simultaneously with the network parameters in a completely unsupervised way, i.e. without requiring access to bead stacks (See pixel maps $\delta_{xy}$  in Fig. \ref{fig:wmap_plots}). We emphasize that neither the training algorithm, nor the network architecture, depends on the specifics of the generative model or the PSF model, and both could well be combined with more flexible functional forms of PSFs.

\subsubsection*{DECODE network architecture for probabilistic single molecule detection and localization}
Our frame analysis module as well as our temporal context module are U-nets with two up- and downsampling stages and 48 filters in the first stage. Each stage consists of three fully convolutional layers, where in each downsampling stage the resolution is halved, and the number of filters doubled, and vice versa in each upsampling stage.
Upsampling is performed using nearest neighbor interpolation to avoid checkerboard artifacts \cite{upsampling}. The final output representation is predicted after two additional convolutions layers.

For each camera pixel, the DECODE network predicts the probability that a fluorophore was detected near that pixel $p$, the location of the detected fluorophore relative to the center of the pixel, $\Delta x,\Delta y,\Delta z$ and the predicted fluorophore brightness $\alpha$, and the uncertainties associated with each of these predictions $\sigma_x,\sigma_y,\sigma_z,\sigma_\alpha$.
We used ELU nonlinear activation function \cite{elu} for all hidden units, the hyperbolic tangent nonlinearity for the coordinate outputs $\Delta x,\Delta y,\Delta z$ and the logistic sigmoid nonlinearity for the non-negative brightness and uncertainty outputs $\alpha,\sigma_x,\sigma_y,\sigma_z,\sigma_\alpha$.

The output channels of the DECODE network together represent a distribution of possible interpretations of a measured image $q(h|d)$ and constitute an approximation to the true posterior distribution over possible detections and localizations $p(h|d)$. Our approximate posterior $q(h|d)$ is a type of Gaussian mixture model, with one Gaussian mixture component per camera pixel. It can represent at most as many particles as there are camera image pixels, and for each pixel represents the detection probability and localization mean and variance of one particle.
In the following sections we describe how this architecture is trained and how deterministic detections and localizations of particles can be obtained from the output of the DECODE network at test-time.

\subsubsection*{Simulator learning}
\label{sec:simlearn}
Given a set of simulated particles and the resulting images over time, simulated as described above such that $h \sim p(h)$ and $d \sim p(d|h)$
we take samples from our generative model $h\sim p(h), d\sim p_\theta(d|h)$ and maximize the loglikelihood $\mathrm{log} q_\phi(h|d)$.
This procedure amounts to minimizing the Kullback-Leibler divergence ($D_{KL}$) between the posterior of the generative model and the recognition network, averaged over the (simulated) data distribution:
\begin{align} \label{eq:simlearn}
\mathbb{E}_{p_\theta(d)} [-D_{KL}(p_\theta(h|d)||q_\phi(h|d)] =
\mathbb{E}_{p_\theta(d,h)} [\mathrm{log} q_\phi(h|d)] + \mathrm{const.}
\end{align}

Given our representation of the final localizations by the discrete pixel positions and the in-pixel offset variables

\begin{align}
x = x_p + \Delta x \quad y = y_p + \Delta y \quad z = \Delta z
\end{align}

we developed a loss function that allows us to jointly optimize the different output variables. We interpret the binary values $p$ as the probability that an activation exists in that pixel while the outputs $\alpha, \Delta x,\Delta y,\Delta z, \sigma_{\alpha}, \sigma_x, \sigma_y,\sigma_z$ parametrize Gaussians $\mathcal{N}(\Vec{\mu},\Sigma)$ which are components of a Gaussian mixture model (GMM) which describes the distribution of emitter activations:

\begin{align}
q_\phi(h|d) \propto \prod_{i}\sum_{k} \frac{p_k}{\sum_{k} p_k} \mathcal{N}(\Vec{X}_i|\Vec{\mu}_k,\Sigma_k)
\end{align}

where $k$ indexes all pixels, and $\Vec{X}_i$ are the ground truth location vectors.

The number of activations follows a Poisson Binomial distribution given that the binary probabilities vary strongly across pixels.
As the likelihood of this distribution is hard to evaluate we instead use its mean and variance to parametrize a Gaussian approximation to the likelihood of counts:

\begin{align}
q_\phi(h|d) \propto \mathcal{N}(\sum_{k} S_k | \sum_{k} p_k, \sum_{k} p_k - p_k^2)
\end{align}

where $\sum_{k}S_k$ is the true number of emitters. 

As the GMM term scales linearly with the number of emitters, while the count term stays constant, we multiply it by the number of emitters to balance the two terms. 
The resulting total log-likelihood we use to train our inference network is:

\begin{align}\label{eq:gmm_loss}
\mathrm{log} q_\phi(h|d) = \sum_{i}\sum_{k} \mathrm{log}\frac{p_k}{\sum_{k} p_k} \mathcal{N}(\Vec{X}_i|\Vec{\mu}_k,\Sigma_k) + \sum_{k}S_k\cdot\mathrm{log}\mathcal{N}(\sum_{k} S_k | \sum_{k} p_k, \sum_{k} p_k - p_k^2)
\end{align}

\paragraph{Auto encoder learning: Optimizing a lower bound on $p(d)$} 
\label{sup:ae_learning}
For auto encoder learning we only treat the discrete outputs as stochastic latent variables, and use the deterministic mean values for the continuous outputs. 
$\mathrm{log} q_\phi(h|d)$ is therefore calculated as the binary cross-entropy between the inferred probabilities and the discrete pixel activations.
We also performed experiments using the loss described in (\ref{eq:gmm_loss}) for autoencoder learning. In this case in addition to the discrete samples we also draw samples of our continuous offsets from $\mathcal{N}(\Vec{\mu}_k,\Sigma_k)$ and optimize them using the reparametrization trick \cite{vae_kingma}. However this resulted in far higher gradient variance and overall reduced performance.  

Using Jensen's inequality we can derive a lower bound (ELBO) on the marginal likelihood $p(d)$:
\begin{align}
\log p(d) = \log \mathbb{E}_{q} [\frac{p_\theta(d,h)}{ q_\phi(h|d)}] \geq 
\mathbb{E}_{q} [\log {\frac{p_\theta(d,h)}{ q_\phi(h|d)}}] = \mathcal{L}(d) 
\label{eq:elbowlong}
\end{align}
by maximizing this ELBO with respect to $\theta$ we minimize the reverse $D_{KL}$ averaged over the true data distribution
\begin{align} 
\mathbb{E}_{p(d)} [D_{KL}(q_\phi(h|d)||p_\mathbf{\theta}(h|d))]
\end{align}
Unlike simulator learning maximization with respect to $\phi$ also allows us to learn the parameters of the generative model. 
If we instead use an importance weighted average over $j$ samples from our recognition model to estimate $p(d)$, and again apply Jensen's inequality we obtain a tighter lower bound (which is identical to the ELBO for $j=1$):
\begin{align}
\mathcal{L}_{IW}^j(d) = \mathbb{E}_{ q(h_{1:J}|d)} \left[ \frac{1}{J}\sum^J_{j=1}\log \bigg[ \underbrace{ \frac{p_\theta(d,h_j)} {q_\mathbf{\phi}(h_j|d)}}_{\omega_k(d,h_j)} \bigg] \right]
\end{align}
This objective is the basis for both the importance weighted autoencoder (IWAE) \cite{iwae} and the reweighted wake-sleep algorithm (RWS) \cite{rws}.

\subsubsection*{Updating $\theta$}

For a given value of $\phi$, unbiased gradients for $\theta$ can be obtained by sampling the discrete variables $h_1,...,h_j \sim q_\phi$ and calculating the gradients:
\begin{align} \label{eq:thetaup}
        \nabla_\theta \mathcal{L}_{IW}^j(d\sim \mathcal{D}) = \nabla_\theta \log (\frac{1}{J}\sum^J_{j=1}\omega_j) = \sum^J_{j=1} \tilde{\omega}_j \nabla_\theta \log p_\theta(h_j|d) \\ \tilde{\omega}_j = \frac{\omega_j(d,h_{j})}{\sum_{j'=1}^J \omega(d,h_{j'})}
\end{align}

\subsubsection*{Updating $\phi$}

Obtaining gradients for $\phi$ is more involved, especially in the case of discrete latents when the reparametrization trick cannot be applied. 

The RWS algorithm includes two procedures to obtain gradients for $\phi$. The sleep phase update matches simulator learning which minimizes the $D_{KL}$ between $p$ and $q$ over data that is generated from the generative model $p_\theta(d|h)p(z)$. 

The wake phase update optimizes the same $D_{KL}$, but over the true data distribution $p(d)$ (i.e. using samples from the data).
\begin{align} \label{eq:rwswake}
\begin{split}
\nabla_\phi D_{KL}(p_\theta(h|d)||q_\phi(h|d)) \simeq \sum^J_{j=1} \tilde{\omega}_j \nabla_\phi \log q_\phi(h_j|d)
\end{split}
\end{align}
We also experimented with the VIMCO algorithm \cite{vimco} and the The Thermodynamic Variational Objective \cite{tvo} as alternative approaches to obtain low variance gradients for discrete latent variable models. Performance was comparable across methods, so we chose RWS for its easy implementation and low number of hyper parameters. 

Wake phase updates can be very noisy, especially during the first iterations when the network basically produces random samples. If the network predicts large numbers of detections training can fail to to memory restraints. Therefore, we start training with a warm up phase of 1000 iterations of simulator learning, where one iteration corresponds to the evaluation of the loss on one batch and a subsequent gradient update.

\subsubsection*{Training details and hyper-parameters}

Training is performed on 40$\times$40 pixel sized regions that are simulated or randomly selected from recorded images at each iteration.
If the network is trained to make use of global context, we use a running average of the hidden states collected over the last 100 training batches. At test time we perform two passes over the dataset: the first one to collect the average hidden state $\tilde{h} = \sum_T h_t$ and the second one to obtain the inference results.

When training with local context we employ different strategies for SL and AE training steps. For simulator learning, when sampling data we align the spatial variables $\Delta x,\Delta y,\Delta z = \vec{\mu}$ (but not the intensity) to be identical when a fluorophore is active in consecutive frames.
For AE training, for each set of variables $S_t, \vec{\mu}_t$ which are inferred from the images $I_{t-1},I_t,I_{t-1}$ we also infer offset variables and uncertainties $\sigma_x, \sigma_y,\sigma_z = \Sigma$ for $t+1$ and $t-1$ to provide context.

We use these variables to calculate an error term that is the sum of log-likelihoods of the offset variables at each pixel under the Gaussian distribution given by the activations in consecutive images:
\begin{align}
    \delta_{xyz} = \sum_{x,y} & S_t\cdot S_{t-1} (\mathcal{N}(\vec{\mu}_t|\vec{\mu}_{t-1},\Sigma_{t-1}) + S_t\cdot S_{t+1}(\mathcal{N}(\vec{\mu}_t|\vec{\mu}_{t+1},\Sigma_{t+1})
\end{align}

This term is subtracted from the objective function during training.

A base value of 0.01 is added to the uncertainties to avoid instabilities.  We used the AdamW optimizer \cite{adamw} with a learning rate of 6$\cdot10^{-4}$ for the network parameters which is multiplied by 0.9 after every 1000 training iterations. When also learning the PSF we used a learning rate of 0.015 for PSF model parameters and 3$\cdot10^{-6}$ for the pixel maps $\delta_{xy}$. To stabilize training and ensure that gradient steps of simulator and autoencoder learning are roughly equal in size when performing combined learning we employ gradient norm clipping with a maximum norm of 0.03. We normalized the inputs to the network by first subtracting the mean of the dataset and then dividing by the maximum of the mean over the image dimension. To calculate the RWS objective we used 40 samples. 

\subsubsection*{Obtaining localizations and post-processing}
\label{sup:nms}

The DECODE network predicts the probabilities of a fluorophore being located at a specific pixel. While we sample from this distribution during training, we prefer to generate deterministic detections and localizations at test time. To get deterministic, fast and precise pseudo samples we instead use a variant of non-maximum suppression to obtain final localizations. 
To obtain a binary mask of fluorophore candidates for a given frame we identify probability peaks, i.e. pixels with values that are above 0.3 and higher than all values in a surrounding 3x3 patch. 
We then add the probability mass from the 4 directly adjacent pixels to the values at the candidate positions by convolving the probability map with a cross shaped filter and applying the mask. All candidates with added probability values above 0.7 are counted towards the localizations. 
The algorithm can be expressed purely in the form of pooling and convolution operations and therefore runs efficiently on a GPU.

For difficult imaging conditions, i.e. high densities, low SNR values and high offsets from the focal plane the lateral offset variables can be biased towards small absolute values. This effect scales with the uncertainty of the predictions and can produce artifacts in the reconstructed image as localizations are concentrated at the pixel centers. 
To counteract this we therefore divide all localizations into equally sized bins according to the total variance  $\mathrm{var}_{tot} = \sqrt{\sigma_{x}^2+\sigma_{y}^2+\sigma_{z}^2}$. Then we calculate an empirical CDF $\hat{F}_{x},\hat{F}_{y}$ from the histograms of the $\Delta x$ and $\Delta y$ variables in each bin. The variables $\Delta\hat{x},\Delta\hat{y} = \hat{F}_{x}(\Delta x)-0.5, \hat{F}_{y}(\Delta y)-0.5$ have a uniform distribution as desired.
This transformation effectively removes image artifacts while having no impact on the performance metrics. 

\begin{figure}[H]
\includegraphics[width=1.0\textwidth,trim={0cm 0.0cm 0.0cm 0.0cm},clip]{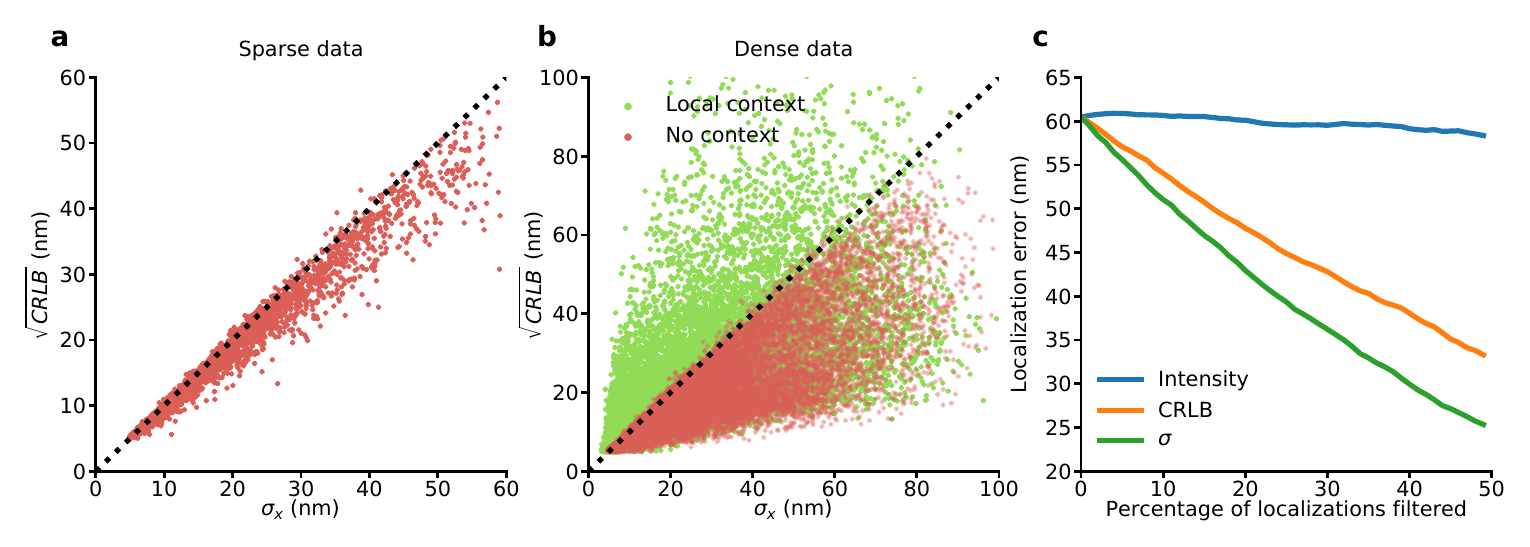}
\caption{{\bf DECODE provides superior uncertainty estimates for dense data} 
CRLB estimates were obtained using the analytical approximation described in \cite{crlb}. {\bf a)} For sparse data with a single emitter per image the uncertainty estimates of DECODE closely match the CRLB. {\bf b)} For dense data DECODE without context produces strictly higher uncertainty estimates, taking into account the reduced precision resulting from overlapping PSFs. When using local context the uncertainty can be lower than the CRLB which doesn't consider temporal dynamics. {\bf c)} For dense data using the $\sigma$ predicted by DECODE to filter out the worst localizations results in lower localization error then using the predicted intensity or the CRLB.
\label{fig:crlb}}
\end{figure}

As a final post-processing step our inferred uncertainties allow us to effectively filter bad localizations. As shown in Fig. \ref{fig:crlb}c this is very effective in reducing the overall localization error. For the challenge data removing between 0 and 20 \% results in the best performance. For real data this threshold should be individually chosen according to the amount of data collected.

\subsubsection*{Evaluating localization accuracy and reconstruction resolution}
To evaluate performance on the challenge datasets, as well as our own simulations we use the lateral or volume localization error in \si{\nano\metre} and the Jaccard index $J$ which quantifies how well an algorithm does at detecting all the fluorophores while avoiding false positives $J = 100\cdot TP/(FN+FP+TP)$. Localizations are matched to ground truth positions when they are withing a circle of 250 \si{nm} radius. As a single metric that evaluates the ability to reliably infer fluorophores with high precision we use the efficiency metric:

\begin{equation}
    E = 100 - \sqrt{(100-J)^2 + \alpha^2 \mathrm{RMSE}^2)}
\end{equation}

Lateral and axial efficiency are calculated with alpha values of $\alpha=0.5 \mathrm{nm}^{-1}$ and $\alpha=1 \mathrm{nm}^{-1}$ respectively and then averaged to obtain the overall 3D efficiency. 
Super-resolution images were rendered by convolving inferred positions with a 2D Gaussian with a width of 5\si{nm}.

The Fourier ring correlation (FRC, \cite{frc,frc_crit}) in \ref{fig:results_ries} was calculated by constructing two super-resolution image volumes of the same sample ($\sigma$=8.5\si{nm}, pixelsize=10\si{nm}) by dividing the localizations into two sets. We did this by alternating blocks of 50k consecutive localizations.

\subsubsection*{DECODE for LLS-PAINT microscopy}
The LLS dataset differs from the the other datasets we analyzed in two respects. First, due to the movement of the light sheet, fluorophores that are active across multiple frames change their $x$ and $z$ position within a frame by a fixed amount. While usually this can be accounted for in post-processing, in order to use local context we also have to adjust our generative model. 
Therefore, when generating the image triplets for simulator learning, we move the emitters by the correct amount when they are active in multiple frames.
Second, the images were recorded with a sCMOS instead of an EMCCD camera, therefore requiring a different noise model.
We follow the description of the image generating process in \cite{scmos} for our noise model. Given the mean intensity as calculated in (\ref{eq:emccd}) it is given by: 

\begin{align}
\label{eq:csmos}
\bar{I}(x,y) =& \sum_{i=1}^N \alpha_i PSF(x-x_i,y-y_i,-z_i) + \beta \\
I(x,y) \sim& \textrm{Gamma}((\bar{I}(x,y))/\eta, \eta)\\
&\eta = \frac{\mathrm{Var(x,y)} + g\cdot(\bar{I}(x,y)-BL)}{\bar{I}(x,y)}.
\end{align}

Here, we assumed the camera gain $g$ to be constant, while $\mathrm{Var(x,y)}$ is the pixel specific noise that can we estimated from dark images. For the results shown in Fig. \ref{fig:lls} we trained DECODE with CL and local context. The PSF was optimized on bead stacks, and the pixel maps $\delta_{x,y}$ where further optimized during the training of the network.

\begin{figure}[H]
\includegraphics[width=1.0\textwidth,trim={0cm 0.0cm 0.0cm 0.0cm},clip]{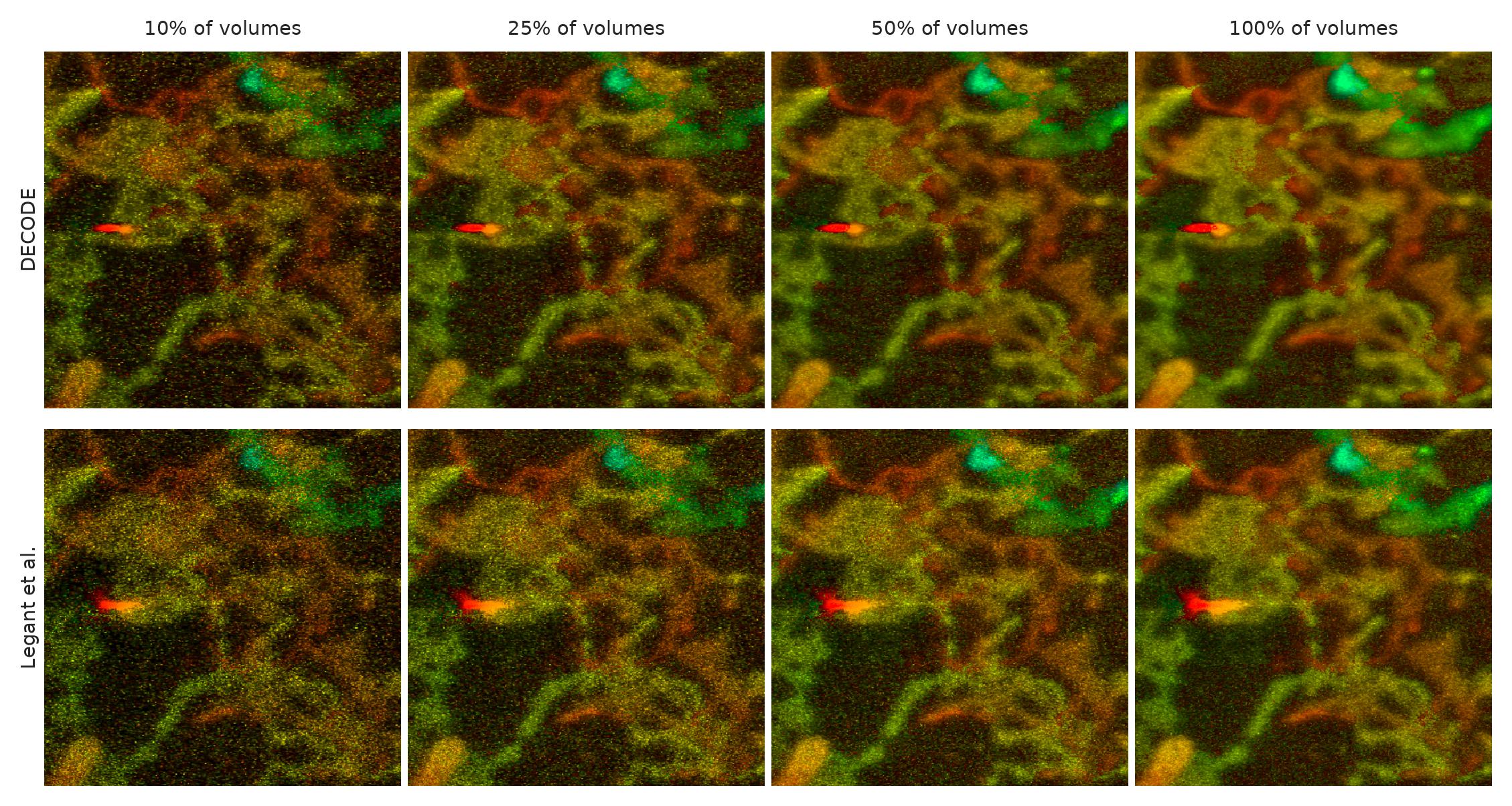}
\caption{{\bf Comparison of reconstructions of LLS data across number of frames} 
Magnified reconstructions (boxed region 5 in Fig. \ref{fig:lls}) using 10, 25, 50 and 100 \% of the available frames.
\label{fig:lls_supp}}
\end{figure}

The reconstructions shown in Figures \ref{fig:lls} and \ref{fig:lls_supp} are rendered by convolving each localization with a 3D Gaussian parametrized by the respective uncertainties to construct a 3D histogram of the volume with a voxel size of 10x10x20 \si{nm}. Histogram values are then clipped at 2.5 to remove most of the contribution from fiducials. We then plotted the color-coded maximum projection over the z-axis with the maximum intensity set to the 99.5 percentile of histogram values.

\vskip 0.2in
\bibliography{refs}

\clearpage

\part*{Supplementary Information}

\beginsupplement
 
\begin{figure}[H]
\includegraphics[width=1.0\textwidth,trim={0cm 0.0cm 0.0cm 0.0cm},clip]{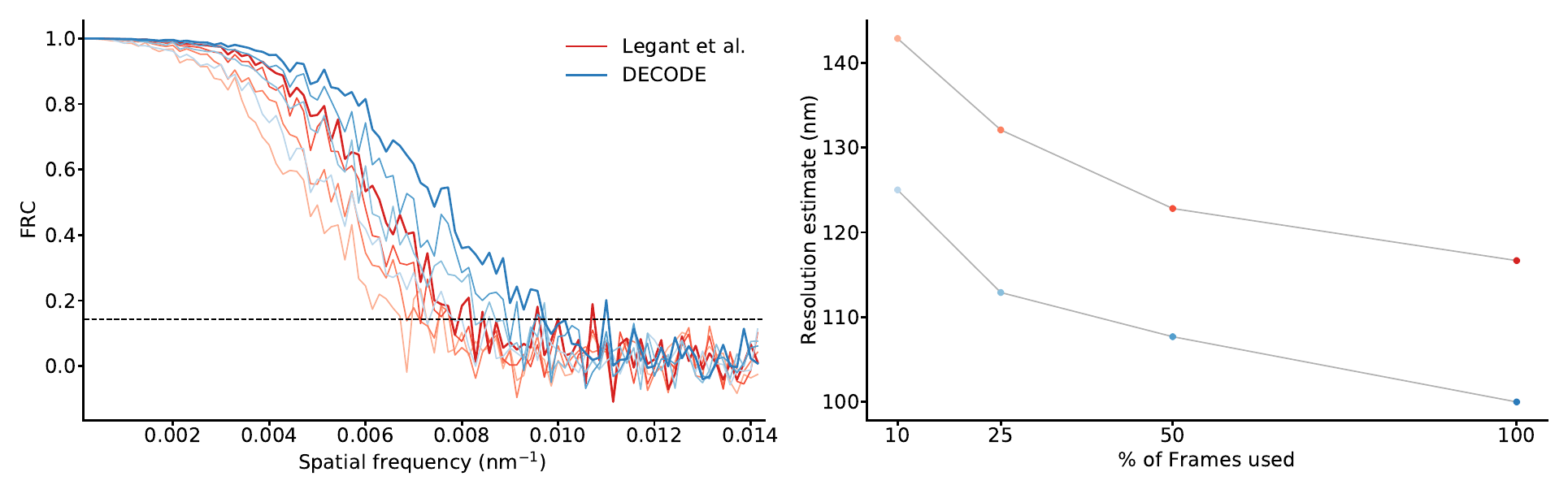}
\caption{{\bf Resolution estimates for LLS reconstructions} 
Resolution estimates obtained using the Fourier Ring Correlation and 0.143 criterion across different percentage of frames used for both methods. Evaluated on the region shown in Fig. \ref{fig:lls_supp}
\label{fig:lls_frc}}
\end{figure}

\end{document}